\documentclass[twocolumn]{article}
\usepackage[letterpaper,margin=1in]{geometry}

\usepackage{amsmath,amssymb}
\usepackage{graphicx}
\usepackage{hyperref}
\usepackage{authblk}
\usepackage{siunitx}
\usepackage{xcolor}
\usepackage{titlesec}
\usepackage{float}
\usepackage[numbers,sort&compress]{natbib}

\usepackage{etoolbox}
\makeatletter
\pretocmd{\subsubsection}{\phantomsection}{}{}
\makeatother

\renewcommand*{\Affilfont}{\normalsize\normalfont}
\renewcommand{\thefootnote}{\fnsymbol{footnote}}

\hypersetup{
    colorlinks,
    linkcolor={red!50!black},
    citecolor={blue!50!black},
    urlcolor={blue!80!black}
}

\newcommand{\unsim}{\mathord{\sim}}
\DeclareSIUnit\angstrom{\text{\AA}}

\newcommand{\revision}[1]{\textcolor{black}{#1}}
\newcommand{\rerevision}[1]{\textcolor{black}{#1}} 

\titleformat{\section}
  {\bfseries\Large}
  {} 
  {0pt} 
  {}
\titlespacing{\section}
  {0pt} 
  {12pt} 
  {0pt} 
\titleformat{\subsection}
  {\bfseries\large}
  {}
  {0pt}
  {}
\titlespacing{\subsection}
  {0pt}
  {6pt}
  {0pt}
\titleformat{\subsubsection}[runin] 
  {\bfseries\normalsize}
  {}
  {0pt}
  {}
\titlespacing{\subsubsection}
  {0pt}
  {6pt}
  {0.3333em}

\usepackage[switch]{lineno}

\title{Crossed laser phase plates for transmission electron microscopy}
\author[1,2]{Petar N. Petrov}
\author[1,2]{Jessie T. Zhang}
\author[1]{Jeremy J. Axelrod\protect\footnotemark[2]}
\author[3]{Pavel K. Olshin}
\author[1,2,*]{Holger M\"uller}

\affil[1]{Department of Physics, University of California, Berkeley, Berkeley, CA 94720, USA}
\affil[2]{Lawrence Berkeley National Laboratory, One Cyclotron Road, Berkeley, CA 94720, USA}
\affil[3]{Biohub, Redwood City, CA 94065, USA}
\affil[*]{Corresponding author: \href{mailto:hm@berkeley.edu}{hm@berkeley.edu}}

\date{\today}

\begin{document}



\twocolumn[
  \maketitle
  \begin{abstract}
  \noindent For decades since the development of phase-contrast optical microscopy, an analogous approach has been sought for maximizing the image contrast of weakly-scattering objects in transmission electron microscopy (TEM). The recent development of the laser phase plate (LPP) has demonstrated that an amplified, focused laser standing wave provides stable, tunable phase shift to the high-energy electron beam, achieving phase-contrast TEM. Building on proof-of-concept experimental demonstrations, this paper explores design improvements tailored to biological imaging. In particular, we introduce the approach of crossed laser phase plates (XLPP): two laser standing waves intersecting in the diffraction plane of the TEM, rather than a single beam as in the current LPP. We provide a theoretical model for the XLPP inside the microscope and use simulations to quantify its effect on image formation. \rerevision{Using simulations, we} find that the XLPP increases information transfer at low spatial frequencies while also suppressing the ghost images formed by Kapitza-Dirac diffraction of the electron beam by the laser beam. We also present a simple acquisition scheme, enabled by the XLPP, which dramatically suppresses unwanted diffraction effects. \rerevision{Finally, we discuss important practical considerations of XLPP design and show experimental results from a prototype.} The results of this study chart the course for future developments of LPP hardware.
  \end{abstract}
  \vspace{1em} 
]

\footnotetext[2]{Current address: Department of Molecular and Cellular Physiology,
Stanford University, Stanford, CA 94305, USA}


\section{Introduction}
\noindent A phase plate increases the contrast of weak-phase objects in transmission electron microscopy (TEM) by phase-shifting the unscattered component of the transmitted electron wave \revision{relative to the scattered component} \cite{Axelrod2024, Glaeser2013, Malac2021}. In the growing field of cryo-electron microscopy (cryo-EM) of biological specimens, such a phase plate is highly sought-after as a means to improve the detection \revision{and alignment} of small proteins \cite{Lander2021, Zhang2020}, discrimination of conformational states of molecules \cite{Herreros2023}, and visualization of multi-scale structural features in electron tomograms \cite{Mahamid2016, Turk2020}. Additionally, a phase plate which affords precise control of the electron beam phase across multiple exposures enables advanced imaging schemes \cite{Petrov2022}.

\revision{While a number of different phase plate designs have been explored in electron microscopy since the first proposals in the 1940s, the implementation of a stable, robust device has proven difficult \cite{Glaeser2013}. Placement of materials in or near the path of the imaging electrons leads to variability in phase plate behavior due to charging and radiation damage to the device, as well as signal loss and imaging artifacts \cite{Danev2014, Obermair2020, Buijsse2020}. While the Volta phase plate, based on a thin carbon film in the diffraction plane of the microscope, has proven the most successful implementation of phase contrast in electron microscopy, the above instabilty and signal loss have discouraged its use \cite{Danev2016, Buijsse2020, Danev2021b}.}

\revision{Recently, the laser phase plate (LPP) demonstrated the potential to improve upon previous designs. By imparting phase shift via Compton scattering, the LPP eschews materials in the path of the electron beam, avoiding the associated pitfalls of material-based phase plates \cite{Axelrod2024}.} The LPP uses a high-intensity continuous-wave laser focus, generated by enhancement in a Fabry-P\'erot cavity, to phase shift the unscattered part of the electron beam \cite{Mueller2010, Schwartz2017, Schwartz2019}. The LPP has demonstrated the full $\frac{\pi}{2}$ \si{\radian} phase shift desired for imaging weak-phase objects \cite{Turnbaugh2021, Danev2011}, as well as stable and analytically-tractable properties over the time needed to take large datasets \cite{Axelrod2020, Axelrod2023}. These features of the LPP have made it a leading candidate to replace previous phase plate designs in cryo-EM \cite{Danev2021, Russo2022}. Application of the LPP to single-particle analysis \revision{has already been demonstrated \cite{Turnbaugh2021, Remis2024}.}

However, a few non-ideal properties of the LPP remain. The relatively large focal radius of the intra-cavity laser beam gives rise to a concomitantly large ``cut-on" spatial frequency, above which phase contrast becomes effective. Compensation for this by magnifying the diffraction pattern leads to increased spherical and chromatic aberration coefficients ($C_s$, $C_c$). Additionally, the LPP generates unwanted ``ghost" images (higher diffraction orders) \cite{Axelrod2020} which, although so weak that they are often buried by noise, may impede its use in the presence of stronger-scattering objects such as the specimen support film or heterogeneous environments such as crowded cellular sections. While increased aberration coefficients may be counteracted by the use of aberration correctors and higher-coherence electron sources, and while partial suppression of ghost artifacts may be achieved by image processing, improvement of the LPP design directly is preferred so that other microscope hardware and software can be used to its full advantage.

\begin{figure*}[htbp!]
    \centering
    \includegraphics{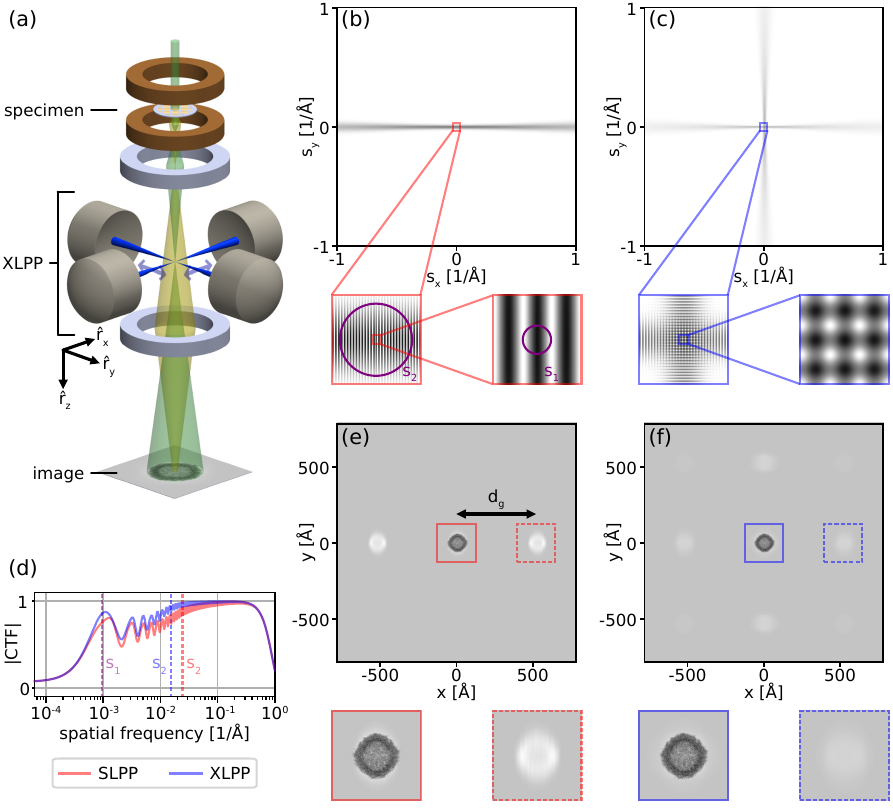}
    \begin{minipage}{0.9\textwidth}
        \caption{\textbf{Crossed laser phase plates (XLPP) concept.} (a) Schematic of a TEM with a XLPP in the conjugate diffraction plane. The incident electron beam (green) is focused at the focus of the XLPP laser beams (blue) while scattered electrons (yellow) are not. \revision{Blue double-headed arrows indicate horizontal laser polarization.} (b-c) Comparisons of the phase shifts $\eta$ produced by the single laser phase plate (SLPP, b) and XLPP (c) as a function of spatial frequency $(s_x,s_y)$. Insets progressively zoom in on lowest spatial frequencies. Purple circles in (b) illustrate the cut-on frequencies $s_2$ (left) and $s_1$ (right). (d) Azimuthally-averaged modulus of the CTF. Vertical dashed lines indicate $s_2$ for the SLPP (red) and XLPP (blue), as well as $s_1$ (purple), which is the same for both. (e-f) Normalized, simulated in-focus images of one apoferritin protein formed using the SLPP (e) and XLPP (f) show ghost images spaced by $d_g$ from the main image. Insets zoom in on main image (solid border) and right ghost image (dashed border). The SLPP has $N_A=0.05$ while the XLPP has $N_A=0.08$. Additional calculation parameters are provided in Table \ref{Table:Parameters}. Color scale ranges from 0 (white) to $\frac{\pi}{2}$ (black) in (b,c) and from 0.08 (black) to 1.1 (white) in (e,f).}
        \label{Fig:Main}
    \end{minipage}
\end{figure*}

Here, we show that combining two LPPs in \revision{a single} diffraction plane at \SI{90}{\degree} to each other in an ``X"-shaped configuration (XLPP, shown schematically in \rerevision{Figures \ref{Fig:Main}a and \ref{Fig:Experiment}a}) can overcome these problems. As we shall see, distributing the laser power and thus the heat load among two cavities can be used to lower the cut-on frequency. Relative to a single LPP (hereafter, SLPP), the XLPP also significantly reduces ghost images and enables novel acquisition schemes which can suppress ghosts further still. The improved imaging properties of the XLPP \revision{are expected to add value for imaging of} large macromolecules and large cellular features such as elements of subcellular ultrastructure \cite{Mahamid2016}. Stronger focusing \revision{or lower laser wavelength} of the LPP will allow $C_s$ and $C_c$ of a phase-plate TEM to be reduced without increasing the cut-on frequency, making the benefits of the LPP more accessible without compensation by advanced, expensive TEM hardware. \revision{In addition, a pair of crossed laser beams with aligned intensity antinodes, in a configuration similar to the prototype XLPP presented here, can be used to correct spherical aberration of an electron microscope, as shown in a recent theoretical study \cite{Uesugi2025}. The XLPP thus introduces new possibilities for TEM, and for coherent electron beam manipulation more generally, which have yet to be fully explored.}

\section{Results} \label{Sec:Results}

\subsection{Electron phase shift} \label{Sec:PhaseShiftTheory}

\noindent The phase shift imparted to an electron by a LPP is calculated following the approach described in \cite{Axelrod2020}. In the case of the XLPP, we consider two linearly-polarized standing waves, propagating in the diffraction plane and at \SI{90}{\degree} to each other (see Figure \ref{Fig:Main}a,c). The total electric field in the diffraction plane is
\begin{linenomath*}
\begin{align}
    &\mathbf{E}(r_x,r_y) = \notag \\
    &\hspace{1mm}2E_0 \Bigg[
        \frac{w_0}{w_x(r_x)} e^{-(r_y^2+r_z^2)/w_x^2(r_x)}
        \cos\!\left(\frac{2\pi}{\lambda_l}r_x\right)\hat{\mathbf{p}}_x \nonumber\\
    &\hspace{5mm}+ \frac{w_0}{w_y(r_y)} e^{-(r_x^2+r_z^2)/w_y^2(r_y)}
        \cos\!\left(\frac{2\pi}{\lambda_l}r_y\right)\hat{\mathbf{p}}_y
\Bigg],
\end{align}
\end{linenomath*}
where $\hat{\mathbf{p}}_{x,y}$ are the polarizations of the two laser beams, $\lambda_l$ is the laser wavelength, $E_0$ is the electric field amplitude, $w_{x,y}$ are the laser beam radii, and  $w_0$ is the minimum radius (``waist"), achieved by each laser beam at their (mutual) focal position. The coordinate system $(r_x,r_y)=(f\lambda_e s_x,f\lambda_e s_y)$ denotes physical space in the diffraction plane, related to the spatial frequency coordinates $(s_x,s_y)$ by the effective focal length $f$ of the microscope and the electron wavelength $\lambda_e$. The physical coordinate along the electron beam axis is denoted $r_z$. In the non-relativistic case, the phase shift of the electron is given by \cite{Mueller2010}
\begin{align}\label{Eq:LineIntegral}
    \eta(r_x,r_y) &= \frac{\alpha\lambda_e}{\hbar\omega^2}\int_{-\infty}^\infty |\mathbf{E}(r_x,r_y,r_z)|^2 \, dr_z
\end{align}
where $\alpha$ is the fine structure constant, $\hbar$ is the reduced Planck constant, $\omega$ is the angular frequency of the laser, and $\lambda_e$ is the electron wavelength. That is, the phase shift $\eta$ is proportional to the integral through the laser intensity along the propagation axis of the electron beam. With the electron beam aligned to the focus and antinode of the SLPP, the unscattered electron beam is phase-shifted by
\begin{align} \label{Eq:CirculatingPowerSLPP}
	\eta_\mathrm{SLPP}(0,0) &= \sqrt{\frac{2}{\pi^3}}\frac{\alpha}{\hbar c^2}\lambda_e\lambda_l N_A P,
\end{align}
where $N_A=\lambda_l/\pi w_0$ is the numerical aperture of the cavity mode and $P$ is the circulating laser power \cite{Schwartz2017}.

\subsubsection{Relativistic effects.}
At the accelerating voltages typically employed in cryo-EM (200-\SI{300}{\kilo\volt}), the relativistic effects in the laser-electron interaction must be taken into account. In general, Equation (\ref{Eq:LineIntegral}) is then no longer valid. For a SLPP, the modulation depth of the phase shift along the standing wave axis becomes polarization-dependent. In the special case of a horizontally-polarized LPP, however, Equations (\ref{Eq:LineIntegral}) and (\ref{Eq:CirculatingPowerSLPP}) remain valid and the maximum phase shift is still achieved at the central antinode \cite{Axelrod2020}.

In the XLPP, the electric fields of the two laser beams add coherently, which will in general give rise to interference effects. In Supplementary Note \ref{Sec:AppendixPolarization}, we show that interfering, vertically-polarized XLPP lasers can sharpen the phase profile for accelerating voltages up to about \SI{100}{\kilo\volt}, and thus lead to a lower cut-on frequency. However, at high accelerating voltages the most favorable phase plate is obtained from two non-interfering, horizontally-polarized laser beams. \revision{This is the special case, illustrated in Figure \ref{Fig:Main}a, which we will consider throughout the main text of this paper.} In this case, Equation (\ref{Eq:CirculatingPowerSLPP}) remains valid, so that the total phase shift of the unscattered beam is simply
\begin{align} \label{Eq:CirculatingPowerXLPP}
	\eta_\mathrm{XLPP}(0,0) &= \sqrt{\frac{2}{\pi^3}}\frac{\alpha}{\hbar c^2}\lambda_e\lambda_l N_A (P_x + P_y),
\end{align}
where $P_{x,y}$ are the circulating powers of the two standing waves. The phase shift profiles created by the SLPP and XLPP are shown in Figure \ref{Fig:Main}b-c.


\subsection{Contrast transfer function} \label{Sec:CTF}

\noindent The structural information contained in the transmitted electron beam is converted into detectable amplitude modulation of the electron beam by the imaging process. Each spatial frequency $\mathbf{s}=(s_x,s_y)$ of electrons scattered in the object plane passes through a point in the diffraction plane. For weak-phase objects such as biomolecules, a suitable mathematical description of the imaging process in the spatial frequency domain is the contrast transfer function (CTF) of the microscope, given by \revision{(Supplementary Note \ref{Sec:AppendixCTFDerivation})}
\begin{align}
    \mathrm{CTF}(\mathbf{s}) = E(\mathbf{s})\sin\left[\chi(\mathbf{s}) - \chi(\mathbf{0}) - \kappa \right]
\end{align}
where $\chi$ is the total phase aberration and $\kappa$ is the contribution from amplitude contrast, which is assumed to be independent of $\mathbf{s}$ \cite{Zivanov2021, Glaeser2021b}. The term $E(\mathbf{s})$ comprises envelope functions which attenuate the CTF (e.g.\ due to partial coherence), but because LPPs do not require any materials to be close to the imaging electrons in the diffraction plane, they do not appreciably attenuate the envelope \cite{Axelrod2023}. The phase shift due to the laser beam(s) adds with the usual phase aberration function (from the microscope lenses) such that
\begin{align} \label{Eq:Chi}
    \chi(\mathbf{s}) &= \frac{2\pi}{\lambda_e}\left( \frac{1}{2}Z\lambda_e^2|\mathbf{s}|^2 + \frac{1}{4}C_s\lambda_e^4|\mathbf{s}|^4 \right) + \eta(\mathbf{s}),
\end{align}
where $Z$ is the defocus and $C_s$ is the coefficient of spherical aberration. For simplicity, other common phase aberrations such as astigmatism and coma, which are typically small in experiment and can be accounted for during data processing, are omitted in this paper.

We note that $\chi(\mathbf{0})=\eta(\mathbf{0})$, the phase shift imparted to the unscattered electron beam solely by the LPP. This means that, in the absence of a phase plate, defocus and spherical aberration are needed to provide appreciable contrast for weak-phase objects. However, in the presence of a phase plate the optimal imaging condition is $Z=C_s=0$. The azimuthal average of the modulus of the CTF is shown in Figure \ref{Fig:Main}d for the SLPP and XLPP in this ``in-focus," (spherical) aberration-corrected imaging configuration. The lack of CTF oscillations at high spatial frequencies when imaging in-focus results in a doubling of the overall spectral power relative to imaging with defocus. \revision{We emphasize that the improvement comes from removing CTF oscillations and not from enhancing the coherence envelope \cite{Fan2017}.} This is illustrated in Supplementary Note \ref{Sec:AppendixCTF} by comparing to a more typical condition of $Z=\SI{-1}{\micro\meter}$ and without $C_s$-correction.

The more closely the phase shift $\eta$ can approximate an ideal Zernike phase plate, which phase-shifts only the unscattered beam at the origin of the diffraction plane, the greater the overall information content of the micrograph (when imaging in-focus). The CTF of a LPP falls short of the ideal CTF of unity in two important ways, which we will now consider.

\subsubsection{Cut-on frequencies.} \label{Sec:CutOnFrequencies}

The finite spatial extent of the antinode of a laser standing wave defines a region in the diffraction plane in which the scattered electron beam receives a similar phase shift to that of the unscattered electron beam, and therefore has a low value of the CTF. This results in two cut-on frequencies characteristic of the LPP, above which the CTF is significantly increased.

The first cut-on frequency is defined as the spatial frequency at which the azimuthal average of $|\mathrm{CTF}(\mathbf{s})|^2$ first reaches 0.5. This approximately corresponds to the lowest spatial frequency which passes through a laser node, which is given by
\begin{align} \label{Eq:FirstCutOn}
    s_1 &= \frac{\lambda_l / 4}{f\lambda_e}.
\end{align}
Although the phase profile of the LPP is not azimuthally-symmetric, Figure \ref{Fig:Main}d demonstrates that contrast increases significantly where $|\mathbf{s}|>s_1$.

A second cut-on frequency of the LPP is defined by the spatial frequency at which the azimuthal average of $|\mathrm{CTF}(\mathbf{s})|^2$ first reaches 0.8. This approximately corresponds to the spatial frequency which is located at a distance from the unscattered beam equal to the waist $w_0=\lambda_l / (\pi N_A)$ of the laser standing wave,
\begin{align} \label{Eq:SecondCutOn}
    s_2 &= \frac{\lambda_l / (\pi N_A)}{f\lambda_e}.
\end{align}

The spatial frequencies $s_1$ and $s_2$ are illustrated in Figure \ref{Fig:Main}b (purple circles). Apart from those which coincide with the streak(s) of laser light in the diffraction plane, spatial frequencies with $|\mathbf{s}|>s_2$ experience no phase modulation and are thus imaged with maximum phase contrast (under the in-focus condition), as seen in Figure \ref{Fig:Main}d. \revision{Weak oscillations of $|\mathrm{CTF}(\mathbf{s})|$ in the region $s_1<|\mathbf{s}|<s_2$ are observed due to the azimuthal average featuring some spatial frequencies which are phase-shifted by the laser light.} Equations (\ref{Eq:FirstCutOn}) and (\ref{Eq:SecondCutOn}) highlight that, theoretically, only four parameters determine the appearance of a laser stripe in the CTF, namely $\{f, \lambda_e, N_A, \lambda_l\}$. While the first two are predetermined by the microscope, we show below that the XLPP supports superior values of $N_A$ and $\lambda_l$ to the SLPP.

\subsubsection{Ghost images.} \label{Sec:GhostImages}

A striking feature of a LPP is the grating-like structure of a laser standing wave, which causes Kapitza-Dirac diffraction of the electron beam \cite{Freimund2001, Schwartz2019}. This diffraction produces ``ghost images" which are spaced by a distance corresponding to
\begin{align}
    d_g = \frac{f\lambda_e}{\lambda_l / 2}
\end{align}
in the specimen plane along the axis of the laser stripe, as illustrated in Figure \ref{Fig:Main}e-f. This distance is inversely proportional to the period of the laser intensity pattern. Although ghost images are faint compared to the main (``zeroth-order diffraction") image, they represent unwanted delocalization of signal over large spatial scales. Similarly to how lower defocus values are sought in standard (defocus-based) cryo-EM to reduce delocalization of signal \cite{Glaeser2021}, large $d_g$ leads to loss of information when ghosts are diffracted beyond the detector and increased noise when ghosts from illuminated objects beyond the field of view diffract onto the detector. In the presence of strong-phase objects or complicated fields of view, ghost images may decrease the interpretability of micrographs.

Further implications of ghost images depend somewhat on the imaging modality. In single-particle cryo-EM, if the main image of one particle is cropped (``picked") out of a micrograph using a box of length $L$, then if $L<2d_g$, the information contained in the particle's ghost images is discarded during data processing. Additionally, ghost images from other nearby particles may be present in the box, resulting in increased background in the box. Requiring instead that $L>2d_g$ retains the delocalized information about the particle but reduces the useful area of the micrograph from which main images can be picked. In either case, ghost images from neighboring particles will average out over a data set and not produce systematic bias in the final reconstruction. On the other hand, when a unique volume is to be reconstructed via tilt series, such as in cryo-electron tomography (cryo-ET), ghost images play a different role. Over the course of a tilt series, objects are expected to move along a circular trajectory about the tilt axis, but ghosts move together with their main images. Failure to account for this will produce artifacts in reconstructed tomograms, but distinguishing main images from ghost images in a large, noisy field of view is nontrivial.

Evidently, suppression of ghosts is a priority in improving imaging with a LPP. It can be accomplished by reducing the intensity of the standing wave, but this must not be done at the expense of sufficient phase shift of the unscattered beam, which is the source of the overall contrast enhancement. It has also been shown that ghost images created by a SLPP can be totally eliminated by setting the laser polarization to the ``relativistic reversal angle," (RRA) but this comes at the expense of substantial loss of low-spatial-frequency contrast since the CTF remains near zero for all spatial frequencies below $s_2$ \cite{Axelrod2020}. In the XLPP, interference of the two laser beams prevents total elimination of the ghost images even when operating at the RRA (Supplementary Note \ref{Sec:AppendixDerivation}). This configuration yields some favorable imaging properties which are explored further in Supplementary Note \ref{Sec:AppendixPolarization}. However, throughout this paper we consider a XLPP with horizontally-polarized laser beams. We demonstrate how this XLPP suppresses ghosts relative to the SLPP without relativistic reversal (Figure \ref{Fig:Ghosts_Spectra}) and enables substantial ghost suppression using a two-image acquisition scheme (Figure \ref{Fig:Deghosting}).


\subsection{Benefits of crossed laser phase plates} \label{Sec:XLPPProperties}

\begin{figure*}[!hptb]
    \centering
    \includegraphics{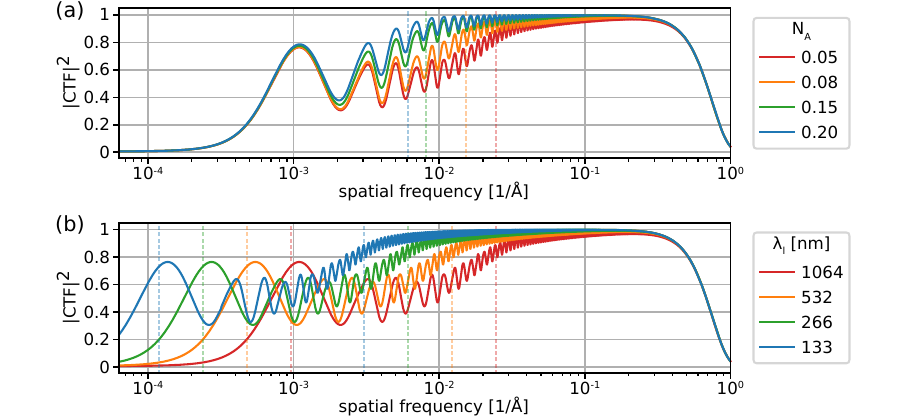}
    \begin{minipage}{0.9\textwidth}
        \caption{\textbf{Improvements of signal power.} Dependence of the square modulus of the CTF, proportional to the power spectral density, on (a) $N_A$ and (b) $\lambda_l$ for a XLPP. In (a), a fixed value of $\lambda_l=\SI{1064}{\nano\meter}$ is used. In (b), a fixed value of $N_A=0.05$ is used. Dashed vertical lines indicate the values of $s_2$ in (a) and both $s_1$ and $s_2$ in (b). Additional calculation parameters are provided in Table \ref{Table:Parameters}. For the condition plotted in red in both panels, $s_1=\num{9.6e-4}/\si{\angstrom}$ and $s_2=\num{2.4e-2}/\si{\angstrom}$.}
        \label{Fig:NA_Wavelength}
    \end{minipage}
\end{figure*}

\subsubsection{Numerical aperture increase.}

As shown in Equation (\ref{Eq:SecondCutOn}), increasing $N_A$ improves the second cut-on frequency, $s_2$. This results in a boost in the signal power at spatial frequencies just below $s_2$, as shown in Figure \ref{Fig:NA_Wavelength}a. The spatial scale $s_2^{-1}\sim\SI{50}{\angstrom}$ for $N_A=0.05$ is comparable to the dimensions of typical cryo-EM targets (e.g.\ apoferritin radius of $\SI{65}{\angstrom}$). As such, information gathered at intermediate spatial frequencies near this value is important for detection of proteins and discrimination of their poses and conformational states \cite{Henderson2011, Scheres2012, Khoshouei2017, Zhong2021}.

Increasing $N_A$ entails focusing the laser more tightly, which is accomplished by bringing the cavity closer to concentricity by 
separating the two cavity mirrors \cite{Schwartz2017}. In the near-concentric limit, $N_A=(2\lambda_l^2/\pi^2R\Delta)^{1/4}$, where $\Delta$ is the distance to concentricity and $R$ is the mirror radius of curvature. As $\Delta$ is decreased, the cavity mode becomes more sensitive to misalignment because its angular deflection upon physical disturbance is magnified. The high light intensity in the cavity also leads to significant heating of the cavity body by scattered light, as well as thermoelastic deformation of the mirror surfaces. These phenomena further exacerbate the effects of misalignments and make locking the cavity more challenging \cite{Turnbaugh2021}. In the SLPP, the $N_A$ has therefore been limited to $\unsim0.05$. With $R=\SI{10}{\milli\meter}$, this entails a distance to concentricity of only $\SI{3.7}{\micro\meter}$. In the XLPP, however, the circulating power required in each cavity to achieve a $\frac{\pi}{2}$ total phase shift is halved (in the non-interfering case considered here), as shown in Equation (\ref{Eq:CirculatingPowerXLPP}). This correspondingly reduces heating of each cavity and thermoelastic deformation of each mirror. We also note that since the phase shift is proportional to $N_A(P_x+P_y)$ (Equation \ref{Eq:CirculatingPowerXLPP}), increasing $N_A$ further reduces the required power in each cavity of the XLPP. Operation of the XLPP at $N_A=0.08$ thus requires only \SI{31}{\percent} of the power in each cavity that is needed for a SLPP with $N_A=0.05$. Preliminary experiments suggest that this reduction in the circulating power should enable an increase of $N_A$ to $\unsim0.08$ in a XLPP.

A comparison between the SLPP and XLPP is shown in Figure \ref{Fig:Main}e-f using simulated images of an isolated protein. The combined effects of lowering the intensity of each laser standing wave and increasing the $N_A$ to 0.08 can be seen. In particular, the contrast of the ghost images is significantly reduced. The enhancement of larger-scale features such as one might see in an electron tomogram of cellular samples is expected to become increasingly pronounced as the spatial scale $s_2^{-1}$ is increased (see Figure \ref{Fig:Main}d).

\begin{figure*}[!hptb]
    \centering
    \includegraphics{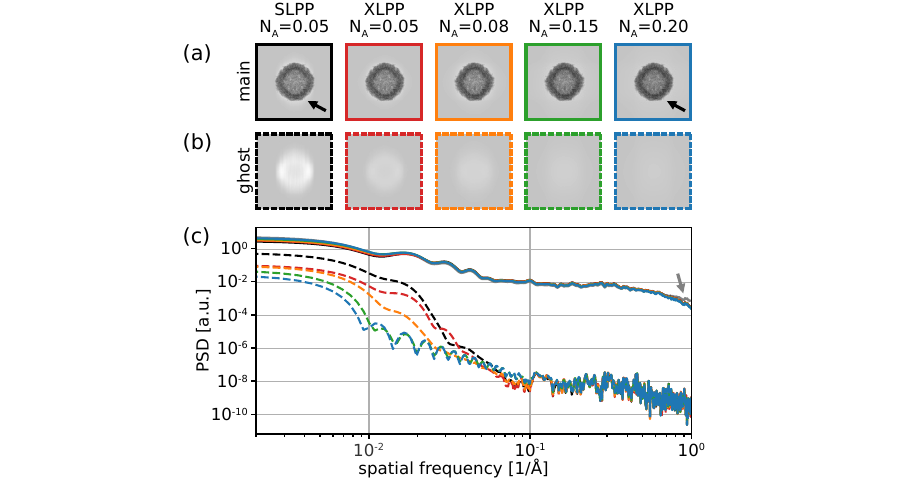}
    \begin{minipage}{0.9\textwidth}
        \caption{\textbf{Ghost suppression by the XLPP.} Simulated noiseless main images (first row) and first-order ghost images (second row) of apoferritin. Color scale ranges from 0.08 (black) to 1.1 (white) in all panels \revision{and each panel side length is \SI{250}{\angstrom}}. Phase plate and $N_A$ are indicated at the top of each column. Arrows point to the light ``halo" around the main image, which is reduced from left to right in the top row. (c) PSDs of main images (solid) and first-order ghost images (dashed). Line colors in (c) correspond to panel border colors in (a) and (b). \revision{PSD of the object is shown in (c) as a gray line indicated by an arrow. It closely follows the other solid lines, departing near 1/(\SI{1}{\angstrom}) due to signal attenuation by the CTF envelope.}}
        \label{Fig:Ghosts_Spectra}
    \end{minipage}
\end{figure*}

Reaching even higher $N_A$ than 0.08 further improves the CTF, as shown in Figure \ref{Fig:NA_Wavelength}a. Figure \ref{Fig:Ghosts_Spectra} also shows that as $N_A$ of the XLPP is increased, ghost images become further suppressed and the ``halo" seen around the main images of particles fades. Reaching the very high values of $N_A$ shown in these figures likely requires modification of the mirror design, which is considered in the Discussion.

\subsubsection{Laser wavelength decrease.} \label{Sec:WavelengthDecrease}

Both cut-on frequencies of the LPP are decreased by decreasing the laser wavelength $\lambda_l$, resulting in significant gains in low-frequency information as shown in Figure \ref{Fig:NA_Wavelength}b. Benefits to the decrease of $s_2$ were summarized in the previous section. Increasing the signal power at very low spatial frequencies improves the contrast of large-scale ($\unsim s_1^{-1}$) features, thereby improving the interpretability of micrographs containing complex biological environments and the contrast of large macromolecular assemblies. This is of especially high interest in cryo-ET \cite{Mahamid2016}.

Decreasing $\lambda_l$ requires some experimental considerations. The scattering loss from mirror surface roughness scales as $\lambda_l^{-2}$ \cite{Bennett1961}, while the power requirement for a LPP scales as $\lambda_l^{-1}$ (see Equation (\ref{Eq:CirculatingPowerSLPP})). Thus, a twofold reduction in $\lambda_l$ results in an eightfold increase in the heat load on a LPP and doubles the required circulating power. Thermoelastic deformation, which increases linearly with circulating power, is then exacerbated. By spreading the light over more cavity mirrors, the XLPP improves the viability of a $\lambda_l$ decrease relative to the SLPP. However, lower-wavelength operation places increasingly stringent demands on the cavity mirrors. In the Discussion, we outline the most important practical aspects deceasing $\lambda_l$.

\subsubsection{Ghost suppression.} \label{Sec:GhostSuppression}

Ghost suppression is a key motivation behind the investigation and development of the XLPP. The use of two lasers, each providing half the total phase shift, suppresses the ghost image artifacts relative to a SLPP, even for a fixed $N_A$. This is illustrated in Figure \ref{Fig:Ghosts_Spectra} using noiseless simulations of an isolated, solvated \cite{Shang2012} protein (apoferritin, PDB 6z6u \cite{Yip2020}) imaged with a SLPP or XLPP \revision{(see Supplementary Note \ref{Sec:AppendixCTFDerivation} and \cite{Petrov2022})}. Due to diffraction along two axes by an XLPP, there are more ghosts in total when using a XLPP, but suppression of their contrast means they will fade further into the background when noise is considered (compare Figures \ref{Fig:Main}e and f). The contrast of first-order ghosts (dashed boxes) is suppressed by the XLPP by a factor of roughly 2-3 relative to the SLPP even when both have $N_A=0.05$, whereas the contrast of the main image is essentially equal between the two cases. Higher-order ghosts are created by both the SLPP and XLPP, but they are very faint and can rarely be observed experimentally, even when imaging strongly-scattering objects such as gold beads. Figure \ref{Fig:Ghosts_Spectra} shows that as $N_A$ is increased, ghost suppression is substantial and occurs in a spatial-frequency-dependent manner. This is because tighter focusing of the laser causes a more rapid decay of light intensity away from the focus and decreases the laser beam waist, $w_0$. The main image is nearly unaffected, although the subtle ``halo" artifact \cite{Remis2024} seen around the protein (indicated by arrows in Figure \ref{Fig:Ghosts_Spectra}) is suppressed by the narrowing of $w_0$.
 
\begin{figure*}[!hptb]
    \centering
    \includegraphics{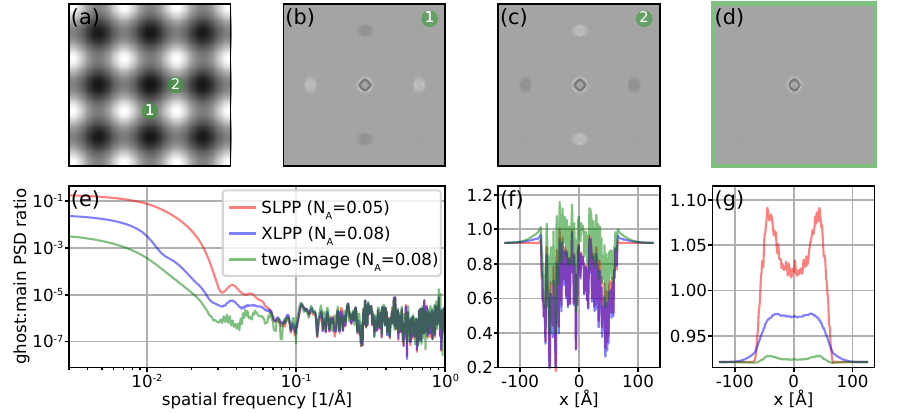}
    \begin{minipage}{0.9\textwidth}
        \caption{\textbf{Two-image scheme for ghost suppression.} (a) Zoomed-in XLPP phase shift $\eta(\mathbf{s})$ showing the location of the unscattered beam for the two images in the two-image sequence, with the color scale ranging zero (white) to $\frac{\pi}{2}$ (black). Panels (b,c) show the first and second image, and (d) shows their average. Color scale in (b-d) is [0.46 (black), 1.06 (white)] \revision{and field of view is \SI{1565}{\angstrom} along each side}. Panel (e) shows the ratio of the power spectral densities of a first-order ghost image and the main image for the case of a SLPP, XLPP, and the two-image result from panel (d). Line scans along the horizontal through the main image (f) and first-order ghost image (g) are shown for the three different cases plotted in (e).}
        \label{Fig:Deghosting}
    \end{minipage}
\end{figure*}

The XLPP enables more sophisticated acquisition schemes that can further suppress ghosts. For example, by slightly shifting the electron diffraction pattern relative to the laser beam between two successive exposures as shown in Figure \ref{Fig:Deghosting}a, two images can be acquired which invert the ghost contrast almost perfectly. The first image is acquired with the unscattered beam in the antinode of one LPP but the node of the other, and the second is acquired in the opposite configuration. After averaging the two images, although the contrast of the main image is slightly lower than in the SLPP and XLPP cases considered (Figure \ref{Fig:Deghosting}f), the ghost contrast is suppressed by a factor of $\unsim20$ relative to the XLPP under normal operation (aligned to the antinode) (Figure \ref{Fig:Deghosting}g). Figure \ref{Fig:Deghosting}e shows the ratio of power spectral densities \revision{(PSDs)} between the ghosts and main images for the three different methods. This two-image approach to ghost suppression only requires deflections of the electron beam by $\unsim$\SI{20}{\micro\radian}, which can be easily implemented with electron beam deflectors, \revision{analogously to the way the SLPP and electron beam are currently aligned in experiments \cite{Remis2024}}. To account for changes in the specimen due to beam-induced motion and radiation damage, the unscattered beam can be switched between the two positions in successive movie frames. \revision{We note that this two-image approach to ghost suppression also works for extended objects (larger than $2d_g$), provided that their phase and amplitude modulation of the electron wave are approximately proportional, as assumed in cryo-EM (see Supplementary Note \ref{Sec:AppendixCTFDerivation}). Furthermore,} this two-image scheme is only a simple case of a much more general class of imaging techniques utilizing phase diversity. Removal of aberrations, as well as recovery of complex (as opposed to real, as typically assumed in cryo-EM) scattering potentials via exit wave reconstruction, are two motivations for developing electron phase-shifting optics like the XLPP. \revision{Such optics like the one described here enable more sophisticated patterning of the electron phase across multiple exposures to extract different types of information about the specimen \cite{Danev2001, Gamm2010, VanDyck2010}.}

\begin{figure*}[!hptb]
    \centering
    \includegraphics{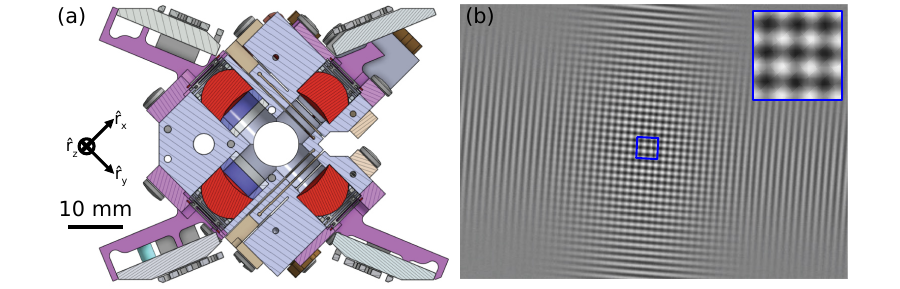}
    \begin{minipage}{0.9\textwidth}
        \caption{\textbf{Prototype XLPP.} (a) Section through a model of a prototype XLPP showing the two laser cavities, each consisting of two mirrors (red), integrated into a single mount. The laser beams (not shown) cross within the bore in the center of the mount. The electron beam (not shown) propagates through the bore, along $\hat{\mathbf{r}}_z$ (into the page). Flat mirrors (gray) steer the two laser beams into and out of the XLPP. (b) Experimental Ronchigram image of the two intersecting standing waves in the XLPP, each with a circulating power of $\unsim$\SI{14}{\kilo\watt}. Inset shows the central region, indicated by the blue box, rotated slightly for comparison to Figure \ref{Fig:Main}c.}
        \label{Fig:Experiment}
    \end{minipage}
\end{figure*}


\rerevision{
\subsection{Prototype crossed laser phase plate} \label{Sec:Experiment}
\noindent A prototype XLPP design that has been constructed at Biohub is shown in Figure \ref{Fig:Experiment}a. Using the successful SLPP design as a starting point \cite{Turnbaugh2021, Axelrod2023}, this prototype uses the same laser wavelength and mirror geometry. As a result, this XLPP occupies a larger volume within the microscope column than the SLPP. To accommodate this, Thermo Fisher Scientific designed several different microscope layouts in which the diffraction plane is accessible via a rectangular port sized \SI{27}{\milli\meter}$\times$\SI{51}{\milli\meter}. This is larger than the port available for previous SLPP work (diameter \SI{25}{\milli\meter}, circular).}

\rerevision{After the XLPP was installed inside the microscope, its two laser beams were visualized in a Ronchigram image \cite{Schwartz2019} shown in Figure \ref{Fig:Experiment}b. By design, the two beams are offset along the electron-optical axis by \SI{100}{\micro\meter} to avoid possible interference effects in this first prototype. Since the laser beams are still approximately within the Rayleigh range of the unscattered beam in the diffraction plane, they appear coplanar and the CTF should be the same as predicted in this paper. Although the cavities share a mount, using two flexures and two laser systems allows the cavities to be stabilized independently. To avoid damaging the cavity mirrors, of which there was a limited supply, this prototype was only tested up to a circulating power of $\unsim$\SI{14}{\kilo\watt} (per cavity), below the circulating power at which we have previously observed laser-induced damage to mirrors in the SLPP. This underscores the value of reducing the required circulating power, which will likely relax manufacturing requirements for the cavity mirrors.}


\section{Discussion} \label{Sec:Conclusions}

\noindent In this paper, we have used simulations and theory to propose and explore a new laser phase plate design based on crossed laser cavities, termed the XLPP. We have shown that the XLPP addresses several limitations of the SLPP. By using a pair of optical cavities, the XLPP enables the laser power to be distributed across two standing waves, thus reducing the heat load on each individual cavity. As a result, the XLPP can provide superior cut-on frequencies by operating at higher numerical aperture or lower laser wavelength. Propagation of the electron beam through a laser standing wave also causes Kapitza-Dirac diffraction, which produces ``ghost" images which delocalize signal electrons over large distances and effectively lead to increased structural noise. We have shown that by allowing lower-power operation of each standing wave, the XLPP suppresses ghost image intensity. We have characterized the behavior of ghost suppression as $N_A$ is increased and also proposed a simple acquisition scheme which further suppresses ghosts, achieving a dramatic reduction of their intensity in composite images. \rerevision{Finally, we have presented a prototype which shows that significant strides toward a XLPP can be made by adapting cavity parameters from the SLPP to a larger mount which is still compatible with realistic microscope designs for high-resolution transmission electron microscopy.}

As discussed above, improvements in $N_A$ and $\lambda_l$ are generally advantageous in terms of the imaging properties of the LPP, and introducing a second cavity in the XLPP affords improvements in either. To realize XLPP operation at very large $N_A$ ($>0.08$) or very small $\lambda_l$ ($<\SI{532}{\nano\meter}$), however, some more adventurous modifications to the XLPP design will be needed.

To enhance $N_A$, the thermoelastic deformation of the cavity mirrors can be further reduced by choosing a different bulk material. Mirrors used in current SLPPs are made of ultra-low expansion glass (ULE Corning Code 7972), the coefficient of thermal expansion (CTE) of which has a zero-crossing just below \SI{300}{\kelvin}. Since the cavity currently operates at a cavity body temperature of $\unsim$\SI{315}{\kelvin}, the related glass ULE Corning Code 7973, which can have a slightly higher CTE, may be a good drop-in replacement. Other candidate materials such as sapphire, silicon, and diamond can be considered for their favorable CTE and thermal conductivity, but their different transmission spectra, operating temperatures near zero CTE, and surface polishing limitations place other constraints on cavity design. As the cavity is brought closer to concentricity, its increasing sensitivity to mechanical misalignment also motivates the development of alternative resonator designs, based on more than two optical components, which are more robust to misalignment \cite{Chen2022, Shadmany2024}. Decreasing the mirror radius of curvature from its current value of $\SI{10}{\milli\meter}$ is another route to higher $N_A$ operation, but this has not yet been achieved for pitch-polished mirrors with smaller $R$ without compromising the mirror surface roughness, which is kept to $<\SI{1}{\angstrom}$ (rms) to minimize scattering losses. Alternative mirror fabrication approaches may be needed \cite{Jin2022}.

In practice, decreasing $\lambda_l$ from its current value of \SI{1064}{\nano\meter} is somewhat more involved than increasing $N_A$. First, a suitable high-power, narrow-linewidth laser with the chosen wavelength must be available \revision{for cavity locking. The SLPP cavity linewidth is $\unsim$\SI{200}{\kilo\hertz}, so a laser with \SI{3}{\kilo\hertz} free-running linewidth is used \cite{Turnbaugh2021}, with a total power of $\unsim$\SI{12}{\watt} at the cavity input.} Second, the increased circulating power linearly increases thermoelastic deformation, so alternative mirror substrates should be considered, as above. At the same time, increased scattering ($\propto\lambda_l^{-2}$) warrants special attention to mirror surface roughness. Finally, contaminants and defects in the mirror coating become increasingly prone to (irreversible) laser-induced damage at higher intensities and lower wavelengths. Low-loss mirror coatings have been reported at \SI{532}{\nano\meter} \cite{Rakhman2016} and defect reduction may be possible \cite{Kong2021}. Fortunately, in our experience mirrors which have once withstood a certain intensity can survive indefinitely in a high-power cavity in the microscope column, so only a small inventory of high-quality mirrors is needed.

\revision{Equations (\ref{Eq:FirstCutOn}) and (\ref{Eq:SecondCutOn}) illustrate that an increase in $N_A$ or decrease in $\lambda_l$ also enables future microscope designs to have shorter effective focal length, $f$, without reducing the cut-on frequency. This bears the advantage that aberration coefficients generally increase with increasing $f$, so the trade-off between de-magnifying the phase plate and increasing $C_c$ and $C_s$ should be evaluated when designing a new phase-contrast TEM instrument. The parameters chosen in this paper (Table \ref{Table:Parameters}) reflect the microscope configuration currently in operation at UC Berkeley.}

In this paper, we have so far assumed that the two laser beams are perfectly co-planar with the diffraction plane and that their relative phase $\Omega$ (see Supplementary Note \ref{Sec:AppendixDerivation}) is tightly controlled. In this important respect, \revision{choosing to operate} the XLPP using horizontally-polarized laser beams simplifies \revision{these design constraints} considerably. Because the horizontally- (and therefore orthogonally-) polarized beams will not interfere, they need not be exactly overlapped along the optical axis, nor have $\Omega$ stabilized. In fact, horizontal displacement of one (or both) of the laser beams by as much as their Rayleigh range, or vertical displacement by $\unsim$\SI{100}{\micro\meter} (comparable to the Rayleigh range of the unscattered electron beam) has only a modest effect on the phase pattern imparted by the LPP, so mechanical alignment tolerances are relaxed compared to the interfering case. \rerevision{This is already observed with the prototype presented here (Figure \ref{Fig:Experiment}b).} A further advantage is that the horizontally-polarized optical field produces a $\frac{\pi}{2}$ \si{\radian} peak phase shift of the electron beam using $\unsim\SI{20}{\percent}$ less circulating laser power (for an accelerating voltage of \SI{300}{\kilo \volt}) than if the lasers are polarized vertically or at the RRA. There are, however, benefits to non-horizontal polarization (see Supplementary Note \ref{Sec:AppendixPolarization}) which may be considered sufficient to undertake the more challenging construction of an XLPP with overlapped and interfering beams in the future.

\rerevision{In summary, in this paper we have shown that, by splitting the laser power between two cavities, the XLPP reduces ghost image intensity and relaxes optomechanical design constraints relative to the SLPP. We have shown how this configuration can be harnessed for further improvements in imaging properties and presented an experimental prototype in a modern microscope.} In addition to advancing the capabilities of phase-contrast cryo-EM as characterized in this paper, development of the XLPP will broaden the parameter space of phase plates available to electron microscopy, enabling advanced imaging schemes as well as providing a wider range of tools to the research community for coherent electron beam manipulation \cite{Danev2001, Gamm2010, Ophus2016, Petrov2022, Axelrod2024}.


\section*{Acknowledgements}
\noindent We thank Bob Glaeser and Osip Schwartz for many useful discussions and gratefully acknowledge our ongoing collaboration with Biohub, as well as Applied Precision Design, LLC, on the construction of a first XLPP -- in particular Anchi Cheng, Yue Yu, Elizabeth Montabana, Noeli Paz Soldan, David Agard, Clinton Potter, Bridget Carragher, Dylan Roof, Matthew Derstine, and Amir Torkaman. We also acknowledge Jake Whinnery (UC Berkeley) for preliminary mechanical design work on the XLPP. We thank Bart Buijsse (Thermo Fisher Scientific) for his design of several relay lens systems for next-generation TEMs compatible with (X)LPPs.

This project was supported by the U.S.\ National Institutes of Health (Grant No.\ 5 R01 GM126011), Chan Zuckerberg Initiative (award number 2021-234606), Gordon and Betty Moore Foundation (Grant No.\ 9366), and a cooperative research and development agreement (CRADA) with Thermo Fisher Scientific (award number AWD00004352). Grant No.\ 5 R01 GM126011 and award number AWD00004352 were administered at Lawrence Berkeley National Laboratory under Contract No.\ DE-AC02-05CH11231. P.N.P. acknowledges support from a postdoctoral fellowship from the National Institute of General Medical Sciences of the National Institutes of Health under Award Number F32GM149186.

\bibliography{references}


\onecolumn
\appendix

\setcounter{secnumdepth}{1}

\renewcommand{\thesection}{\arabic{section}}  
\titleformat{\section}
  {\normalfont\Large\bfseries} 
  {Supplementary Note~\thesection:} 
  {0.5em} 
  {\hangindent=0pt} 

\renewcommand{\theequation}{S\arabic{equation}}
\setcounter{equation}{0}
\renewcommand{\thefigure}{S\arabic{figure}}
\setcounter{figure}{0}
\renewcommand{\thetable}{S\arabic{table}}
\setcounter{table}{0}


\section{Derivation of the relativistic XLPP phase shift} \label{Sec:AppendixDerivation}

\noindent In the non-relativistic limit, the interaction between the electromagnetic field and the electron beam is given by the ponderomotive potential, which does not depend on the polarization of the electromagnetic field and repels the electrons from the high-intensity antinodes of the laser standing wave. For relativistic electron beams, as found in practice, the interaction becomes polarization-dependent. For electrons moving at velocities $v_e$ larger than $c/\sqrt{2}$ (electron energies larger than $\unsim$\SI{211}{\kilo e\volt}), the interaction can even become reversed, such that the antinodes appear attractive to the electron beam while the nodes become repulsive \cite{Axelrod2020}. In this section, we will derive the relativistic phase shift profile created by the XLPP.

In the laboratory frame in Coulomb gauge, the phase shift $\eta$ imparted to the electron matter waves by the laser is \cite{Axelrod2020}
\begin{align} \label{Eq:EtaGeneral}
    \eta &= \frac{\mathrm{e}^2}{2m\hbar}\int \frac{dt}{\gamma}\left[ \left(\mathbf{A}(\mathbf{d}_0(t),t)-\nabla G(\mathbf{d}_0(t),t)\right)^2 -\beta^2 \left(A_z(\mathbf{d}_0(t),t)-\nabla_z G(\mathbf{d}_0(t),t)\right)^2 \right],
\end{align}
with $-\mathrm{e},m$ the electron charge and mass, $\mathbf{A}$ the vector potential of the laser field with $A_z:=\mathbf{A}\cdot\hat{\mathbf{r}}_z$, $\mathbf d_0(t)$ the unperturbed electron trajectory, $G$ a gauge function that arises from the Lorentz transformation between the lab frame and the frame in which the uperturbed electron beam is at rest, $\beta = v_e/c$ the normalized electron velocity, and $\gamma=1/\sqrt{1-\beta^2}$. The vector potential $\mathbf{A}(\mathbf{r},t) = \mathbf{A}_1(\mathbf{r},t) + \mathbf{A}_2(\mathbf{r},t)$ of the XLPP is given by adding two standing waves propagating in the $\hat{\mathbf{r}}_x$- and $\hat{\mathbf{r}}_y$-direction, respectively:
\begin{align}
    \mathbf{A}_1(\mathbf{r},t) &= A_1(y,z)\cos(kx) \left[\cos(\theta_1)\cos(\omega t)\hat{\mathbf{r}}_z + \sin(\theta_1)\cos(\omega t - \varepsilon_1)\hat{\mathbf{r}}_y\right], \\
    \mathbf{A}_2(\mathbf{r},t) &= A_2(x,z)\cos(ky) \left[\cos(\theta_2)\cos(\omega t + \Omega)\hat{\mathbf{r}}_z + \sin(\theta_2)\cos(\omega t - \varepsilon_2 + \Omega)\hat{\mathbf{r}}_x\right]
\end{align}
where $\mathbf{r}:=(r_x,r_y,r_z)$ are the spatial coordinates, $\theta_j$ are polarization angles relative to the $\hat{\mathbf{r}}_z$ axis, $\varepsilon_j$ are ellipticity parameters, $k=2\pi/\lambda_l$ is the angular wave number, and $\Omega$ is the temporal phase of the two standing waves. Assuming the envelope functions $A_{1,2}$ are slowly-varying relative to the wave cycle along the electron trajectory \cite{Axelrod2020}, we can time-average the integrand of Equation (\ref{Eq:EtaGeneral}) over one period $T=2\pi/\omega$ of the optical field, writing $\eta$ in terms of an effective potential $U$,
\begin{align} \label{Eq:EtaEffectivePotential}
    \eta &= \frac{1}{\hbar} \int dt\, U(\mathbf{r}), \\
    U(\mathbf{r}) &= \frac{\mathrm{e}^2}{2m\gamma}\frac{1}{T}\int_0^T dt\, \left[ \left(\mathbf{A}(\mathbf{r},t)-\nabla G(\mathbf{r},t)\right)^2 -\beta^2 \left(A_z(\mathbf{r},t)-\nabla_z G(\mathbf{r},t)\right)^2 \right],
\end{align}
where we note that $r_z=c\beta t$. In the slowly-varying envelope approximation, we can also approximate the gauge function as \cite{Axelrod2020}
\begin{align}
    G(\mathbf{r},t) &\approx \frac{c\beta}{\omega} \left[ A_1(r_y, r_z) \cos(kr_x) \cos(\theta_1) \sin(\omega t) + A_2(r_x, r_z) \cos(kr_y) \cos(\theta_2) \sin(\omega t + \Omega) \right].
\end{align}
Under these conditions, Equation (\ref{Eq:EtaEffectivePotential}) gives the phase shift for an arbitrary configuration of the parameters $\{\theta_1,\theta_2,\varepsilon_1,\varepsilon_2,\Omega\}$.

\subsection{Special cases}
\noindent We will consider the three special cases of laser beams polarized vertically, horizontally, and at the ``relativistic reversal angle."

\subsubsection{Vertical polarization.}
When the two laser polarizations are vertical ($\theta_{1,2}=0$), we find
\begin{align} \label{Eq:UVertical}
    U(\mathbf{r}) = U_v(\mathbf{r}) &= \frac{\mathrm{e}^2}{4m\gamma} \Bigg\{ \frac{A_1^2(r_y,r_z)}{2}\left[1 + (1-2\beta^2)\cos(2kr_x)\right] + \frac{A_2^2(r_x,r_z)}{2}\left[1 + (1-2\beta^2)\cos(2kr_y)\right] \notag \\
    &\hspace{25mm} + 2A_1(r_y,r_z)A_2(r_x,r_z)\cos(kr_x)\cos(kr_y)\cos(\Omega)(1-\beta^2) \Bigg\}.
\end{align}
We note that the terms in $A_j^2$ have the same form as those in a single phase plate \cite{Axelrod2020}, wherein the effective potential appears as a standing wave with its modulation depth scaled by $(1-2\beta^2)$. In the non-relativistic limit $\beta\to0$, we observe that the potential reduces to
\begin{align} \label{Eq:UVerticalNonrelativistic}
    U_v(\mathbf{r}) &= \frac{\mathrm{e}^2}{4m\gamma} \lvert A_1(r_y,r_z)\cos(kr_x) + A_2(r_x,r_z)\cos(kr_y)e^{i\Omega} \rvert^2,
\end{align}
such that the phase shift $\eta$ is proportional to the integral of the laser intensity along $\hat{\mathbf{r}}_z$. However, we note that at the accelerating voltages used in cryo-EM, the $\beta$-dependent terms cannot be neglected. The consequences of this for the CTF are explored in Supplementary Note \ref{Sec:AppendixPolarization}.

\subsubsection{Horizontal polarization.}
When both laser polarizations are horizontal ($\theta_{1,2}=\frac{\pi}{2}$), the two laser beams can no longer interfere and the effective potential is simply the sum of that coming from two SLPPs,
\begin{align} \label{Eq:UHorizontal}
    U(\mathbf{r}) = U_h(\mathbf{r}) &= \frac{\mathrm{e}^2}{4m\gamma} \Bigg\{ \frac{A_1^2(r_y,r_z)}{2}\left[1 + \cos(2kr_x)\right] + \frac{A_2^2(r_x,r_z)}{2}\left[1 + \cos(2kr_y)\right] \Bigg\}.
\end{align}
We can see that the resulting potential is similar to that of Equation (\ref{Eq:UVertical}) in the limit $\beta\to1$ except that the modulation depth of the standing waves is inverted in the latter case.

\subsubsection{Relativistic reversal angle polarization.}
\begin{figure*}[!hptb]
    \centering
    \includegraphics{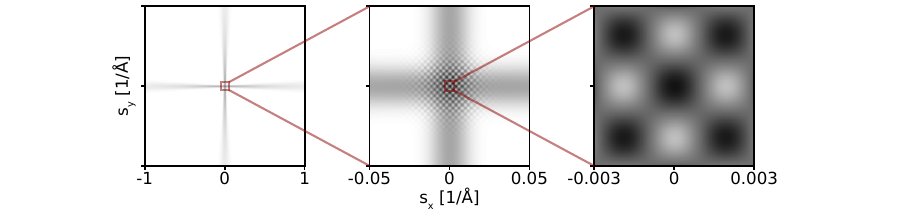}
    \begin{minipage}{0.9\textwidth}
        \caption{\textbf{Phase shift at the relativistic reversal angle.} Progressively zoomed-in view of the phase shift produced by a XLPP with both laser beams polarized at the RRA (with $N_A=0.08$). Middle panel shows the difference between the central region where the beams interfere and the outer regions where the standing wave structure of the laser intensities is washed out by the polarization-dependent relativistic effect. White corresponds to zero phase shift and black to the maximum.}
        \label{Fig:EtaRRA}
    \end{minipage}
\end{figure*}

When $\beta\geq 1/\sqrt{2}$, the polarization-dependence of the laser-electron interaction permits a unique feature of $\eta$. At the so-called relativistic reversal angle (RRA), $\theta_r:=\arccos(1/\sqrt{2}\beta)$, the nodes and antinodes of a laser standing wave create the same phase shift \cite{Axelrod2020}. In this case, the potential of the XLPP simplifies to
\begin{align} \label{Eq:UReversal}
    U(\mathbf{r}) = U_r(\mathbf{r}) &= \frac{\mathrm{e}^2}{4m\gamma} \Bigg\{ \frac{1}{2}\left[A_1(r_y,r_z)^2 + A_2(r_x,r_z)^2\right] + \frac{1-\beta^2}{\beta^2}A_1(r_y,r_z)A_2(r_x,r_z)\cos(\Omega)\cos(kx)\cos(ky) \Bigg\}.
\end{align}
The phase shift away from the origin takes on a smooth Gaussian shape, while near the origin the pattern is complicated by the interference of the two laser beams. The phase pattern that results is plotted in Figure \ref{Fig:EtaRRA}.

\section{Effects of laser polarization} \label{Sec:AppendixPolarization}

\noindent As shown in Supplementary Note \ref{Sec:AppendixDerivation}, the phase shift $\eta$ produced by the XLPP depends significantly on the polarization of the laser beams. In this section, we consider the effects of laser polarization on the CTF by examining the three special cases derived above: when the lasers are polarized vertically, horizontally, and at the RRA.

\begin{figure*}[!hptb]
    \centering
    \includegraphics{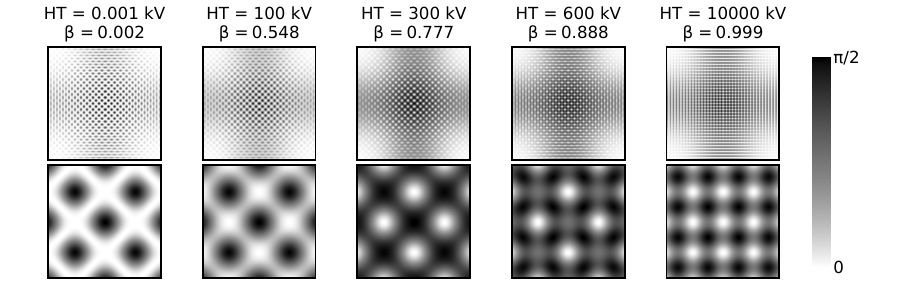}
    \begin{minipage}{0.9\textwidth}
        \caption{\textbf{Dependence of phase shift $\eta$ on electron velocity $\beta$ in a vertically-polarized XLPP.} Each column shows the phase shift $\eta$ as experienced by electrons with a different accelerating voltage (HT) through a XLPP with vertically-polarized ($\theta_{1,2}=0$) lasers. Bottom panel is zoomed in 10$\times$ around the center relative to the top panel and in all cases the phase shift is linearly scaled so that the peak phase shift is $\frac{\pi}{2}$ (black). \revision{Top panel field of view is \SI{20}{\micro\meter} while bottom panel field of view is \SI{2}{\micro\meter} along each side.}}
        \label{Fig:Relativistic}
    \end{minipage}
\end{figure*}

Vertical polarization maximizes the interference of the two lasers. The cross-term produced by laser interference results in a favorable cut-on frequency in the non-relativistic case (Equation \ref{Eq:UVerticalNonrelativistic}), but examination of the relativistic interaction reveals that this feature gradually disappears and the cut-on frequency gradually increases as the electron velocity $\beta$ is increased (Figure \ref{Fig:Relativistic}). It should be noted that $\lambda_e$ changes with $\beta$, which affects the size of the diffraction pattern.

\begin{figure*}[!hptb]
    \centering
    \includegraphics{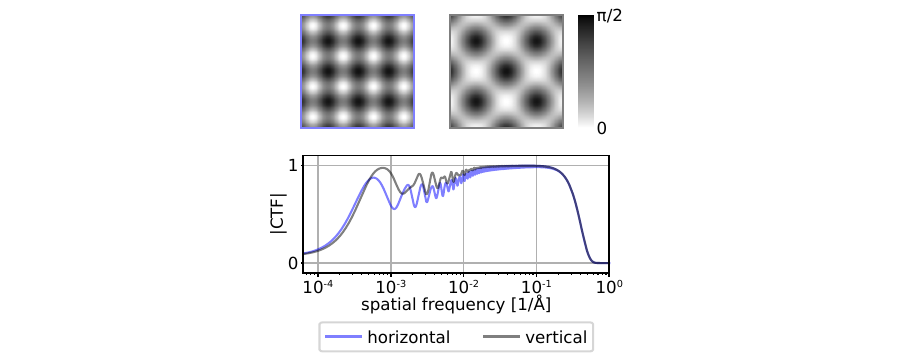}
    \begin{minipage}{0.9\textwidth}
        \caption{\textbf{Polarization-dependence of $\eta$ at \SI{100}{\kilo\volt} accelerating voltage.} Top panel shows the zoomed-in phase shift for horizontally-polarized beams (left) and vertically-polarized beams (right)\revision{, each with a field of view of \SI{2}{\micro\meter}}. In both cases, laser power is set so that peak phase shift is $\frac{\pi}{2}$. Bottom panel shows the azimuthally-averaged modulus of the CTF.}
        \label{Fig:100kV}
    \end{minipage}
\end{figure*}

When the lasers are horizontally-polarized, they cannot interfere, so the cut-on frequency remains fixed as $\beta$ is varied. For low $\beta$, therefore, this configuration produces a worse cut-on frequency than vertical polarization, but as seen in Figure \ref{Fig:Relativistic}, the performance of the latter is limited at high $\beta$. The transition occurs around an accelerating voltage of $\unsim\SI{100}{\kilo\volt}$, as shown in Figure \ref{Fig:100kV}.

\begin{figure*}[!hptb]
    \centering
    \includegraphics{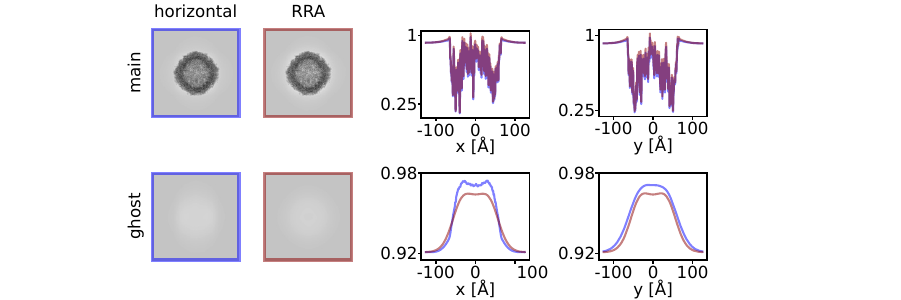}
    \begin{minipage}{0.9\textwidth}
        \caption{\textbf{Ghost suppression by relativistic reversal.} Top row shows a main image of an isolated apoferritin particle for a XLPP with $N_A=0.08$ and either horizontal (blue) or RRA (red) polarization. Horizontal (x) and vertical (y) line scans through the centers of the two images are shown. Bottom row shows the same plots for first-order ghost images. RRA polarization suppresses peak ghost contrast by $\unsim\SI{20}{\percent}$. Horizontal scans show that high-spatial-frequency information in the ghost is lower with RRA polarization than horizontal polarization. \revision{Field of view in images is \SI{250}{\angstrom} along each side.}}
        \label{Fig:GhostsRRA}
    \end{minipage}
\end{figure*}

When the accelerating voltage exceeds $\unsim\SI{211}{\kilo \volt}$, operation of the lasers at the RRA is a viable option to eliminate the modulation of $\eta$ along the laser beams at high spatial frequencies, as seen in Figure \ref{Fig:EtaRRA}. Relative to horizontal polarization, the ghost image contrast is suppressed by $\unsim\SI{20}{\percent}$ in this configuration, and the smoothing of $\eta$ at high spatial frequencies suppresses the high-spatial-frequency components in the ghost images (Figure \ref{Fig:GhostsRRA}). However, the cut-on frequency is somewhat worse.

\begin{figure*}[!hptb]
    \centering
    \includegraphics{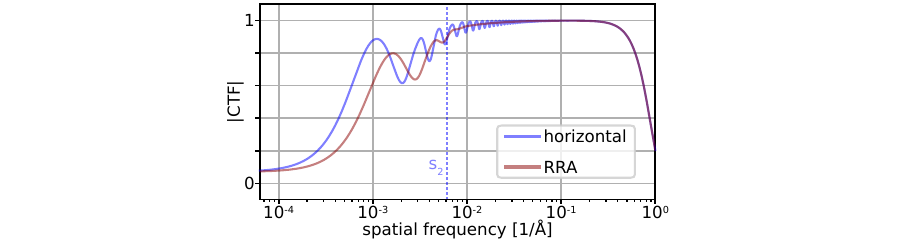}
    \begin{minipage}{0.9\textwidth}
        \caption{\textbf{Contrast transfer at $N_A$ of 0.2.} Azimuthally-averaged modulus of the CTF is plotted for horizontal and RRA polarizations of the XLPP. The former has a lower first cut-on frequency but the latter does not have amplitude oscillations above $s_2$ due to the washing out of the standing wave patterns away from the laser intersection.}
        \label{Fig:CTFRRA}
    \end{minipage}
\end{figure*}

Evidently, vertical polarization is favored at accelerating voltages at or below $\unsim\SI{100}{\kilo \volt}$ for its low cut-on frequency, while horizontal polarization is favorable at higher voltages. At voltages above $\unsim\SI{211}{\kilo \volt}$, operation at the RRA may be desirable for its suppression of ghosts. However, the reduction in ghost contrast is modest and the amplitude of high-spatial-frequency components of ghosts is relatively low (Figure \ref{Fig:Ghosts_Spectra}), so it remains to be seen whether this configuration is worth the considerably more demanding alignment than the horizontal XLPP. Figure \ref{Fig:CTFRRA} compares the azimuthally-averaged modulus of the CTF between horizontal and RRA polarizations of the XLPP at the very high $N_A$ of 0.2, demonstrating that, despite a slight increase in cut-on frequency, the latter yields contrast transfer satisfyingly close to the ideal Zernike phase plate.

\section{Removal of CTF oscillations} \label{Sec:AppendixCTF}

\begin{figure*}[!hptb]
    \centering
    \includegraphics{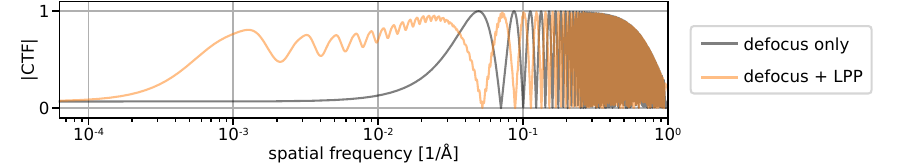}
    \begin{minipage}{0.9\textwidth}
        \caption{\textbf{CTF oscillations caused by defocus and spherical aberration.} Azimuthally-averaged modulus of the CTF is plotted for the case of $Z=\SI{-1}{\micro\meter}$ and $C_s=\SI{2.7}{\milli\meter}$ in the absence of a phase plate (gray) and when using a SLPP with $N_A=0.05$ (orange). Additional calculation parameters are provided in Table \ref{Table:Parameters}.}
        \label{Fig:DefocusCTF}
    \end{minipage}
\end{figure*}

\noindent As discussed above, the use of a phase plate in cryo-EM obviates the need for defocus and spherical aberration. Operating with a SLPP or XLPP in the ``in-focus" condition with spherical aberration correction ($Z=C_s=0$) yields the CTFs shown in Figure \ref{Fig:Main}d, which have no oscillations at high spatial frequencies. This results in a doubling of the spectral power of images at the high spatial frequencies relative to the usual case of $Z\approx\SI{-1}{\micro\meter}$ and $C_s\approx\SI{2.7}{\milli\meter}$. For comparison, the latter configuration is illustrated in Figure \ref{Fig:DefocusCTF}, both with and without a LPP. While incorporating a phase plate into this configuration increases the CTF substantially at low spatial frequencies ($|\mathbf{s}|<s_2$), the phase plate alone is not sufficient to eliminate the oscillations of the CTF at high spatial frequencies, which are caused by the terms in $\chi$ (Equation (\ref{Eq:Chi})) which are quadratic (defocus) and quartic (spherical aberration) in $|\mathbf{s}|$.

\section{Calculation parameters}
\noindent Table \ref{Table:Parameters} lists the typical values of relevant parameters used in the calculations throughout this study. Deviations from these values are specified in the text, figure captions, and/or figure legends.

\begin{table}[H]
\centering
\begin{tabular}{|c|c|c|}
    \hline
    Parameter & Symbol & Value \\
    \hline
    \revision{effective} focal length & $f$ & \SI{14.1}{\milli\meter} \\
    energy spread (FWHM) & -- & \SI{0.3}{e\volt} \\
    amplitude contrast & $\kappa$ & 0.07 \\
    specimen thickness & -- & \SI{250}{\angstrom} \\
    chromatic aberration & $C_c$ & \SI{5.1}{\milli\meter} \\
    spherical aberration & $C_s$ & \SI{2.7}{\milli\meter} or zero \\
    defocus & $Z$ & \SI{-1}{\micro\meter} or zero \\
    electron wavelength & $\lambda_e$ & \SI{1.97}{\pico\meter} \\
    mirror radius of curavture & $R$ & \SI{10}{\milli\meter} \\
    laser wavelength & $\lambda_l$ & \SI{1064}{\nano\meter} \\
    laser polarization ellipticities & $\varepsilon_{1,2}$ & zero \\
    laser polarization angles & $\theta_{1,2}$ & $\pi/2$ \si{\radian} \\
    laser temporal phase & $\Omega$ & zero \\
    \hline
\end{tabular}
\begin{minipage}{0.9\textwidth}
\caption{Values of parameters used in calculations throughout the paper, unless otherwise indicated. Parameters indicated with -- are not assigned a symbol in the text. The non-zero values of $Z,C_s$ are used in the ``conventional" configuration in Figure \ref{Fig:DefocusCTF}; elsewhere, the ``in-focus" condition is used, with $Z=C_s=0$.}
\label{Table:Parameters}
\end{minipage}
\end{table}


\revision{\section{Derivation of the CTF}} \label{Sec:AppendixCTFDerivation}
\revision{\noindent To derive the CTF as used in this paper, we consider image formation in the case of a plane wave of electrons traveling along the optical axis and incident upon a weakly, elastically scattering specimen. The ``exit wave" $\psi_e$ may be written
\begin{align}
    \psi_e(\mathbf{x}) &= e^{i\varphi(\mathbf{x})}e^{-\mu(\mathbf{x})} = 1 + \psi_s(\mathbf{x})
\end{align}
where $\mathbf{x}=(x,y)$ denotes the spatial coordinates in the object (specimen) plane of the microscope and $\psi_s$ denotes the ``scattered" component of the electron wave function. The microscope is described by the transfer function $H(\mathbf{s}) = E(\mathbf{s})\cdot e^{-i\chi(\mathbf{s})}$, where $E(\mathbf{s})$ describes the coherence envelope and $\chi(\mathbf{s})$ is the phase aberration in Equation (\ref{Eq:Chi}). The image formed on the detector is given by
\begin{align} \label{Eq:ImageFormation}
    \lvert \psi_i(\mathbf{x}) \rvert^2 &= \lvert\mathcal{F}^{-1}[\mathcal{F}[\psi_e]\cdot H](\mathbf{x})\rvert^2 \\
    &= \lvert H(\mathbf{0}) + \left(\psi_s*\mathcal{F}^{-1}[H]\right)(\mathbf{x}) \rvert^2
\end{align}
with $\mathcal{F}$ denoting the two-dimensional Fourier transform and $\mathcal{F}^{-1}$ its inverse. Simulations of apoferritin in this paper use Equation (\ref{Eq:ImageFormation}) to form images from the exit wave $\psi_e$. The contribution of inelastically-scattered electrons is neglected throughout, i.e.\ use of an energy filter is assumed.
To leading order in $\psi_s$ (i.e.\ in the weak scattering case), we have
\begin{align}
    \lvert \psi_i(\mathbf{x}) \rvert^2 &= |H(\mathbf{0})|^2 + H(\mathbf{0})\cdot\left(\psi_s*\mathcal{F}^{-1}[H]\right)^*(\mathbf{x}) + H^*(\mathbf{0})\cdot\left(\psi_s*\mathcal{F}^{-1}[H]\right)(\mathbf{x})
\end{align}
with $(\cdot)^*$ denoting complex conjugation. The Fourier transform of the image is then
\begin{align}
    \mathcal{F}[\lvert \psi_i \rvert^2](\mathbf{s}) &= |H(\mathbf{0})|^2\delta(\mathbf{s}) + \Psi_s^*(-\mathbf{s})H(\mathbf{0})H^*(-\mathbf{s}) + \Psi_s(\mathbf{s})H^*(\mathbf{0})H(\mathbf{s}),
\end{align}
with $\Psi_s:=\mathcal{F}[\psi_s]$. In the weak scattering case, $\psi_s(\mathbf{x}) \approx i\varphi(\mathbf{x}) - \mu(\mathbf{x})$ with $\varphi(\mathbf{x}),\mu(\mathbf{x})\in\mathbb{R}$, so
\begin{align}
    \Psi_s^*(-\mathbf{s}) &= -i\Phi^*(-\mathbf{s}) - M^*(-\mathbf{s}) = -i\Phi(\mathbf{s}) - M(\mathbf{s})
\end{align}
where $\Phi:=\mathcal{F}[\varphi]$ and $M:=\mathcal{F}[\mu]$. We can then write
\begin{align} \label{Eq:ImageFT}
    \mathcal{F}[\lvert \psi_i \rvert^2](\mathbf{s}) &= |H(\mathbf{0})|^2\delta(\mathbf{s}) + 2\Phi(\mathbf{s})\left\{\frac{i}{2}\left[ H^*(\mathbf{0})H(\mathbf{s}) - H^*(-\mathbf{s})H(\mathbf{0})\right]\right\} \notag \\
    &\hspace{6mm}- 2M(\mathbf{s})\left\{\frac{1}{2}\left[ H^*(\mathbf{0})H(\mathbf{s}) + H^*(-\mathbf{s})H(\mathbf{0})\right]\right\}
\end{align}
The term in the braces which multiplies $\Phi$ is the (phase) contrast transfer function. To further specialize and recover Equation (\ref{Eq:Chi}), we make some common assumptions.}

\revision{First, we assume that the transfer function is symmetric to inversion, $H(\mathbf{s})=H(-\mathbf{s})$. Then Equation (\ref{Eq:ImageFT}) becomes
\begin{align}
    \mathcal{F}[\lvert \psi_i \rvert^2](\mathbf{s}) &= E^2(\mathbf{0})\delta(\mathbf{s}) + 2\Phi(\mathbf{s})E(\mathbf{0})E(\mathbf{s})\sin\left[\chi(\mathbf{s}) - \chi(\mathbf{0})\right] - 2M(\mathbf{s})E(\mathbf{0})E(\mathbf{s})\cos\left[\chi(\mathbf{s}) - \chi(\mathbf{0})\right].
\end{align}}

\revision{Second, we make the approximation, widely adopted in cryo-EM, that amplitude modulation of the scattered wave is proportional to its phase modulation, $\mu=\kappa\varphi$ for $\kappa\ll1$ such that
\begin{align}
    \mathcal{F}[\lvert \psi_i \rvert^2](\mathbf{s}) &= E^2(\mathbf{0})\delta(\mathbf{s}) + 2\Phi(\mathbf{s})E(\mathbf{0})E(\mathbf{s})\left\{\sin\left[\chi(\mathbf{s}) - \chi(\mathbf{0})\right] - \kappa\cos\left[\chi(\mathbf{s}) - \chi(\mathbf{0})\right]\right\} \\
    &\approx E^2(\mathbf{0})\delta(\mathbf{s}) + 2\Phi(\mathbf{s})E(\mathbf{0})E(\mathbf{s})\sin\left[\chi(\mathbf{s}) - \chi(\mathbf{0}) - \kappa\right].
\end{align}
With the straightforward assumption that $E(\mathbf{0})=1$, we arrive at
\begin{align}
    \mathcal{F}[\lvert \psi_i \rvert^2](\mathbf{s}) &= \delta(\mathbf{s}) + 2\Phi(\mathbf{s})\cdot\mathrm{CTF}(\mathbf{s}), \\
    \mathrm{CTF}(\mathbf{s}) &:= E(\mathbf{s})\sin\left[\chi(\mathbf{s}) - \chi(\mathbf{0}) - \kappa\right].
\end{align}
In the absence of a phase plate, $\chi(\mathbf{0})=0$ is often set by convention and, as such, that term is sometimes not made explicit elsewhere in the literature. However, this expression illustrates that phase contrast is determined by the phase which is accumulated in relation to the unscattered wave at the origin of the diffraction plane, $\mathbf{s}=\mathbf{0}$.}

\end{document}